\documentclass[twoside,twocolumn,9pt]{article}
\usepackage{extsizes}
\usepackage[super,sort&compress,comma]{natbib} 
\usepackage[version=3]{mhchem}
\usepackage[left=1.5cm, right=1.5cm, top=1.785cm, bottom=2.0cm]{geometry}
\usepackage{balance}
\usepackage{widetext}
\usepackage{times,mathptmx}
\usepackage{sectsty}
\usepackage{graphicx} 
\usepackage{lastpage}
\usepackage[format=plain,justification=raggedright,singlelinecheck=false,font={stretch=1.125,small,sf},labelfont=bf,labelsep=space]{caption}
\usepackage{float}
\usepackage{fancyhdr}
\usepackage{fnpos}
\usepackage[english]{babel}
\usepackage{array}
\usepackage{droidsans}
\usepackage{charter}
\usepackage[T1]{fontenc}
\usepackage[usenames,dvipsnames]{xcolor}
\usepackage{setspace}
\usepackage[compact]{titlesec}

\usepackage{dcolumn}   
\usepackage{bm}        
\usepackage{amsmath}   
\usepackage{amsfonts}
\usepackage{amssymb}
\usepackage{multirow,bigdelim}
\usepackage{hyperref}
\hypersetup{
   colorlinks=true,       
   linkcolor=black,         
   citecolor=blue,        
   urlcolor=blue           
}

\usepackage{epstopdf}

\definecolor{cream}{RGB}{222,217,201}

\begin{document}

\pagestyle{fancy}
\thispagestyle{plain}
\fancypagestyle{plain}{

\fancyhead[C]{\includegraphics[width=18.5cm]{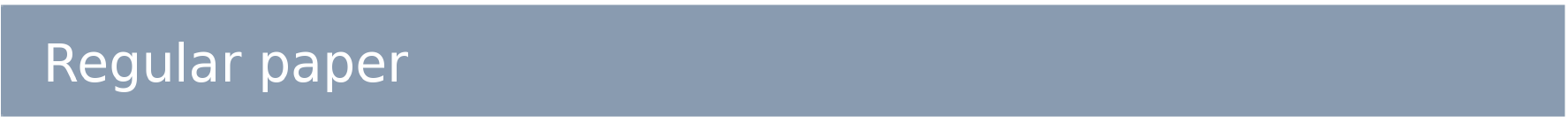}}
\fancyhead[L]{\hspace{0cm}\vspace{1.5cm}\includegraphics[height=30pt]{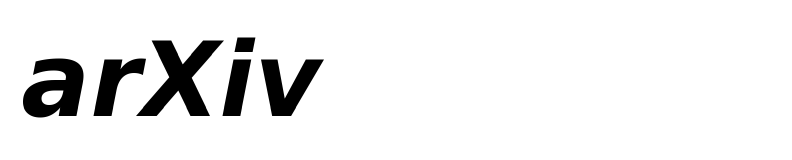}}
\renewcommand{\headrulewidth}{0pt}
}

\makeFNbottom
\makeatletter
\renewcommand\LARGE{\@setfontsize\LARGE{15pt}{17}}
\renewcommand\Large{\@setfontsize\Large{12pt}{14}}
\renewcommand\large{\@setfontsize\large{10pt}{12}}
\renewcommand\footnotesize{\@setfontsize\footnotesize{7pt}{10}}
\makeatother

\renewcommand{\thefootnote}{\fnsymbol{footnote}}
\renewcommand\footnoterule{\vspace*{1pt}%
\color{cream}\hrule width 3.5in height 0.4pt \color{black}\vspace*{5pt}} 
\setcounter{secnumdepth}{5}

\makeatletter 
\renewcommand\@biblabel[1]{#1}            
\renewcommand\@makefntext[1]%
{\noindent\makebox[0pt][r]{\@thefnmark\,}#1}
\makeatother 
\renewcommand{\figurename}{\small{Fig.}~}
\sectionfont{\sffamily\Large}
\subsectionfont{\normalsize}
\subsubsectionfont{\bf}
\setstretch{1.125} 
\setlength{\skip\footins}{0.8cm}
\setlength{\footnotesep}{0.25cm}
\setlength{\jot}{10pt}
\titlespacing*{\section}{0pt}{4pt}{4pt}
\titlespacing*{\subsection}{0pt}{15pt}{1pt}

\fancyfoot{}
\fancyfoot[RO]{\footnotesize{\sffamily{1--\pageref{LastPage} ~\textbar  \hspace{2pt}\thepage}}}
\fancyfoot[LE]{\footnotesize{\sffamily{\thepage~\textbar\hspace{3.45cm} 1--\pageref{LastPage}}}}
\fancyhead{}
\renewcommand{\headrulewidth}{0pt} 
\renewcommand{\footrulewidth}{0pt}
\setlength{\arrayrulewidth}{1pt}
\setlength{\columnsep}{6.5mm}
\setlength\bibsep{1pt}

\makeatletter 
\newlength{\figrulesep} 
\setlength{\figrulesep}{0.5\textfloatsep} 

\newcommand{\topfigrule}{\vspace*{-1pt}%
\noindent{\color{cream}\rule[-\figrulesep]{\columnwidth}{1.5pt}} }

\newcommand{\botfigrule}{\vspace*{-2pt}%
\noindent{\color{cream}\rule[\figrulesep]{\columnwidth}{1.5pt}} }

\newcommand{\dblfigrule}{\vspace*{-1pt}%
\noindent{\color{cream}\rule[-\figrulesep]{\textwidth}{1.5pt}} }

\hyphenation{ALPGEN}
\hyphenation{EVTGEN}
\hyphenation{PYTHIA}

\renewcommand{\d}{\partial}
\renewcommand{\div}{\mathrm{div}\,}
\newcommand{\dd}{\mathrm{d}}
\newcommand{\eps}{\varepsilon}
\newcommand{\deps}{\dot\varepsilon}
\renewcommand{\L}{\mathrm{L}}
\newcommand{\W}{\mathrm{W}}
\newcommand{\loc}{\mathrm{loc}}
\newcommand{\dev}{\mathrm{dev}}
\renewcommand{\H}{\mathrm{H}}
\newcommand{\R}{\mathbf{R}}
\newcommand{\C}{\mathcal{C}}
\newcommand{\T}{\mathcal{T}}
\newcommand{\D}{\mathcal{D}}
\newcommand{\A}{\mathcal{A}}
\renewcommand{\S}{\mathcal{S}}
\newcommand{\sph}{\mathfrak{S}}
\renewcommand{\O}{\mathcal{O}}
\newcommand{\V}{\mathcal{V}}
\newcommand{\Q}{\mathcal{Q}}
\newcommand{\F}{\mathcal{F}}
\newcommand{\tendsto}{\rightarrow}
\newcommand{\Norm}[1]{\|#1\|}
\newcommand{\lap}{\Delta}
\newcommand{\gammap}{\dot{\gamma}}
\newcommand{\p}{\mathbf{P}}
\renewcommand{\u}{\mathbf{u}}
\newcommand{\elemp}{\mathbb{P}}
\newcommand{\ex}{\mathbf{e}_x}
\newcommand{\ey}{\mathbf{e}_y}
\newcommand{\ez}{\mathbf{e}_z}
\newcommand{\Heb}{H\'ebraud{}}
\newcommand{\rem}{\mathrm{rem}}
\newcommand{\Rb}{\overline{R}}
\newcommand{\Qb}{\overline{Q}}
\newcommand{\cb}{\overline{c}}
\newcommand{\ct}{\tilde c}
\newcommand{\Gt}{\widetilde G}
\newcommand{\Gb}{\overline{G}}
\newcommand{\Sh}{\widehat S}
\newcommand{\conv}{\ast}
\renewcommand{\equiv}{\Leftrightarrow}
\newcommand{\Meas}{\mathcal{M}}
\newcommand{\app}{\mathrm{app}}
\newcommand{\B}{\mathcal{B}}
\newcommand{\I}{\mathcal{I}}
\newcommand{\J}{\mathcal{J}}
\newcommand{\valabs}[1]{\vert #1\vert}

\newtheorem{prop}{Proposition}
\newtheorem{theo}{Theorem}
\newtheorem{cor}{Corollary}
\newtheorem{lemma}{Lemma}

\let\oldhat\hat
\renewcommand{\vec}[1]{\mathbf{#1}}
\renewcommand{\hat}[1]{\oldhat{\mathbf{#1}}}

\makeatother

\twocolumn[
  \begin{@twocolumnfalse}
\vspace{3cm}
\sffamily
\begin{tabular}{m{4.5cm} p{13.5cm} }

 & \noindent\LARGE{\textbf{Probing relevant ingredients in mean-field approaches for the athermal rheology of yield stress materials}} \\
\vspace{0.3cm} & \vspace{0.3cm} \\

 & \noindent\large{Francesco Puosi,$^{\ast}$\textit{$^{a,b,c}$} Julien Olivier,\textit{$^{d}$} and Kirsten Martens\textit{$^{b,c}$}} \\

 & \noindent\normalsize{	}

\end{tabular}

 \end{@twocolumnfalse} \vspace{0.6cm}

  ]

\renewcommand*\rmdefault{bch}\normalfont\upshape
\rmfamily
\section*{}
\vspace{-1cm}

\footnotetext{\textit{$^{\ast}$~Corresponding author: francesco.puosi@ens-lyon.fr}}
\footnotetext{\textit{$^{a}$~Laboratoire de Physique de l'\'Ecole Normale Sup\'erieure de Lyon, Universit\'e de Lyon, CNRS, 46 All\'ee d'Italie, 69364 Lyon c\'edex 07, France.}}
\footnotetext{\textit{$^{b}$~Universit\'e Grenoble Alpes, LIPHY, F-38000 Grenoble, France.}}
\footnotetext{\textit{$^{c}$~CNRS, LIPHY, F-38000 Grenoble, France.}}
\footnotetext{\textit{$^{d}$~Aix Marseille Universit\'e, Centre Math\'ematiques et Informatique, Techn\^opole Ch\^ateau Gombert, 39 Rue F Joliot Curie, F-13453 Marseille 13, France.}}


\section{Introduction}
The theoretical understanding of the yielding transition in athermally driven disordered systems is a highly challenging problem and no consensus has been established even on the basic ingredients that should underlie coarse grained descriptions of the non-linear rheological response of yield stress materials \cite{ SGR, HL, FalkPRE57, RodneyMSME11, NicolasEPL14, AgoritsasEPJE15}. 

The only commonly accepted point of view is that disordered materials, such as glasses or soft matter, exhibit a strongly heterogeneous dynamics when driven by an external shear. Fast 
particle rearrangements, the so-called shear transformations, take place in small regions while the rest of the material deforms elastically \cite{Argon79}. These
plastic events induce long-range elastic deformations in the system leading to complex correlations of the yielding regions in form of plastic avalanches \cite{Eshelby57, PuosiPRE14,LemaitrePRL09, MartensPRL11, SalernoPRE13, BudrikisPRE13, LinPNAS14}. 

One of the major concerns remains to know whether 
despite this dynamical complexity 
it is sensible to describe the yielding dynamics within mean-field descriptions \cite{SGR, FalkPRE57, HL, KEP1, DahmenNatPhys11}. We tackle this key question using particle based simulations, concentrating on flow responses in a regime that is relevant for many rheological setups \cite{CantatJPCM05, BecuPRL06, GoyonNature08, GoyonSoftMat10,AmonPRL12,KamrinPRL12,DesmondSoftMat13, KnowltonSoftMat14, LeBouilPRL14, JensenPRE14, GendelmanPRE2014}. We find that at high enough shear rates and/or small enough system sizes we recover a dynamics well described by mean-field considerations \cite{LiuArXiv15}, similar to near mean-field critical points in equilibrium phase transitions \cite{Kac63, KleinPRE07}.

\begin{figure}[htpb]
\begin{center}
\includegraphics[width=0.8\columnwidth, clip]{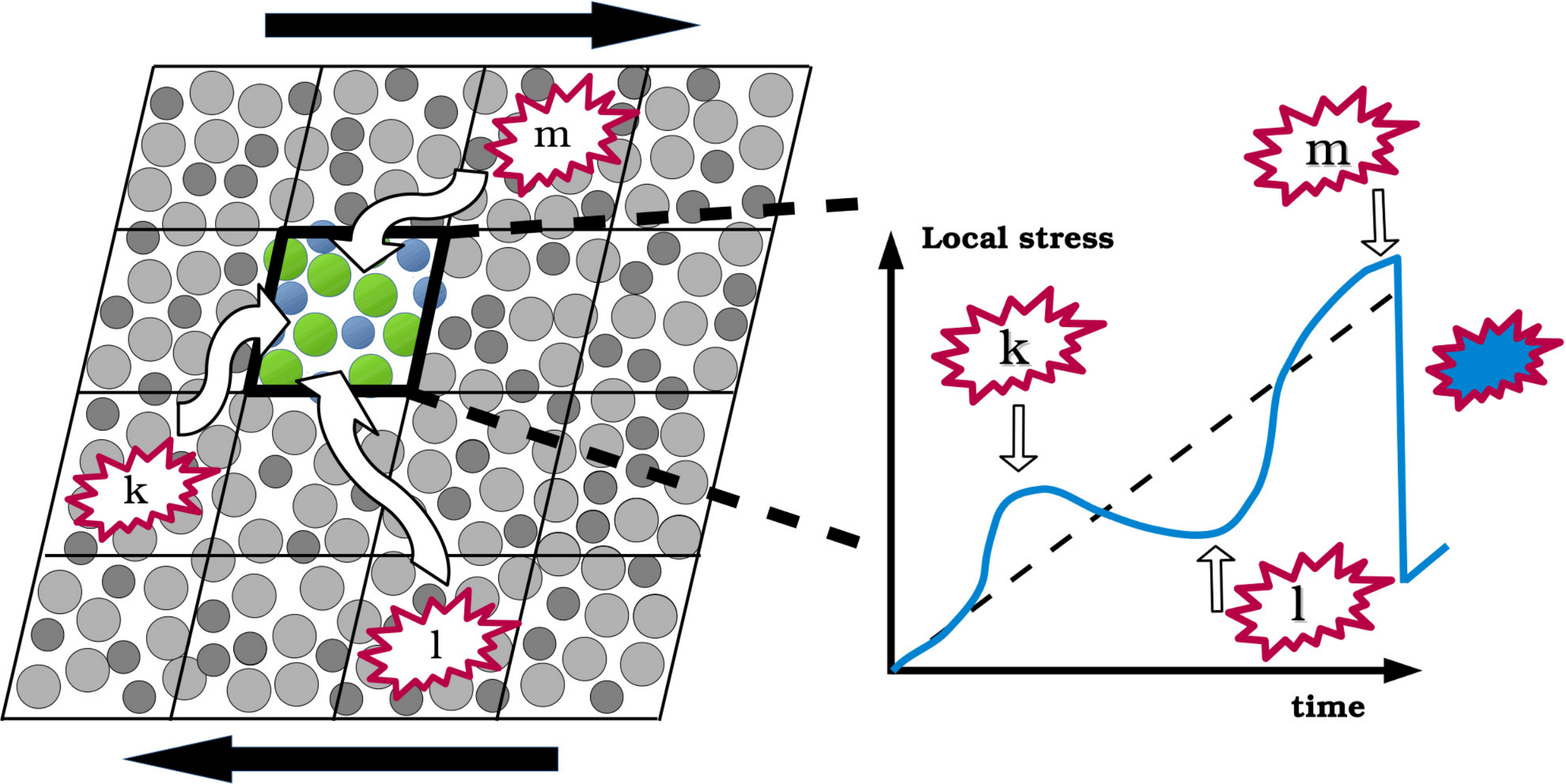}
\end{center}
\caption{{\it Schematic view of the coarse grained picture:} The particle simulation box is devided into smaller mesoscopic parts that can hold exactly one plastic event. The plastic activity in the surrounding of a mesoscopic region (see graphical interpretation on the right) leads to stress fluctuations around the mean value of the stress (dashed line in the graph on the right), which is controlled by the external forcing.
}
\label{fig:figure1}
\end{figure}

An important feature that found its way into several theories, e.g.~the Soft Glassy Rheology (SGR) and the Shear Transformation Zone Theory (STZ), is to describe yielding in a generalized thermodynamic description with an effective temperature \cite{SGR, FalkPRE57}. But the analogy with thermally activated events proposed by the SGR theory has recently been questioned for athermal rheology \cite{NicolasEPL14, AgoritsasEPJE15}. Also it has been shown that the thermodynamic interpretation of the effective temperature is problematic in the athermal regime \cite{XuPRL05, PuckettPRL13} and that the STZ theory is not able to predict the non-linear response \cite{LangerArXiv15}, expected in the small driving limit \cite{LiuArXiv15}.

In this work we show that the dynamically created noise in sheared amorphous systems can be encompassed through a normal diffusion in the local stresses \cite{Karmakar-PRE2010} (for a schematic view see Fig.~\ref{fig:figure1}). In agreement with former works \cite{NicolasEPL14, AgoritsasEPJE15}, we consider that local stress diffusion is acting as noise in the tilt of the local potential energy landscape with an amplitude proportional to the plastic activity, an observable that can be measured experimentally \cite{AmonPRL12, KnowltonSoftMat14}. 

We analyse within molecular dynamics simulations the coupling strength between the mechanical noise and the plastic activity. Within our simulations we successfully relate this coupling strength, a dimensionless and density independent quantity, in a consistent manner to the flow response without any further parameters. This finding leads us to reconsider a speculative mean-field scenario put forward a long time ago in the so-called H\'ebraud-Lequeux (HL) model \cite{HL}, that was at the basis of important further developments \cite{KEP1, KEP2, KamrinPRL12}, providing so far one of the best self-consistent mean-field description of mechanical noise in athermally sheared disordered systems.

\section{Mean-field approach}
\label{sec:theo}
Within the framework of the HL-model we can establish a link between the dynamical yield stress and the prefactor 
of a Herschel Bulkley type power law fit of the rheological curves of athermally sheared yield stress materials. 
This relation should, according to this theory, solely be determined by the coupling strength between mechanical noise and the rate of activity.
In the following we review briefly the main assumptions and results of the mean-field approach \cite{HL}.
The model describes the state of a soft glassy material via
the probability density $\mathcal{P}$ of local shear stresses $\sigma=\sigma_{xy}$ in regions of mesoscopic size $W$
while the material is sheared at rate $\gammap$. The time evolution of $\mathcal{P}$ is given by
\begin{eqnarray}\label{eq:HLmodel}
\partial_t \mathcal{P}(\sigma ,t)
 =	&	- G_0 \dot{\gamma}(t)  \partial_\sigma \mathcal{P}
 		- \frac{1}{\tau} \theta (\valabs{\sigma} - \sigma_c)  \mathcal{P} \nonumber\\
 	& 	+ \Gamma (t) \delta (\sigma)
 	        + D_{\text{HL}}(t) \partial_{\sigma}^2 \mathcal{P} 		
\end{eqnarray}	
where $\theta(x)$ and $\delta(x)$ denote respectively the usual Heaviside and delta-distributions. 
The first term on the right hand side proportional to the stress gradient of the probability density $\partial_\sigma \mathcal{P}$ accounts for the linear elastic response.
The following term describes the loss in the probability density due to local yielding of overstressed regions above a critical stress $\sigma_c$ at a rate given by $1/\tau$.
It has been argued that the phenomenological parameter $\tau$ can be interpreted as the duration of a plastic event in the low shear rate limit \cite{AgoritsasEPJE15}. The corresponding gain term is given in the third expression on the right hand side, where the stress is set to zero after a yielding event with a rate given by the plastic activity rate 
\begin{equation}\label{eq:actrate}	
 \Gamma(t)=\frac{1}{\tau}\int_{|\sigma|>\sigma_c} \mathcal{P}(\sigma,t)\dd\sigma\;.
\end{equation}

The last term encompasses the stress changes created through other yielding events
in a mean-field manner, assuming that this mechanical noise can be approximated through a normal diffusion of the mesoscopic stresses.
To describe this noise in an self-consistent manner, the HL approach proposes that its amplitude should be related to 
the rate of plastic activity through a dimensionless coupling constant $\tilde{\alpha}$
\begin{equation}\label{eq:DHL}
 D_{\text{HL}}(t) = \tilde{\alpha} \sigma_c^2 \Gamma(t)\;.
\end{equation}

\begin{figure}[th]
\begin{center}
\includegraphics[width=\columnwidth, clip]{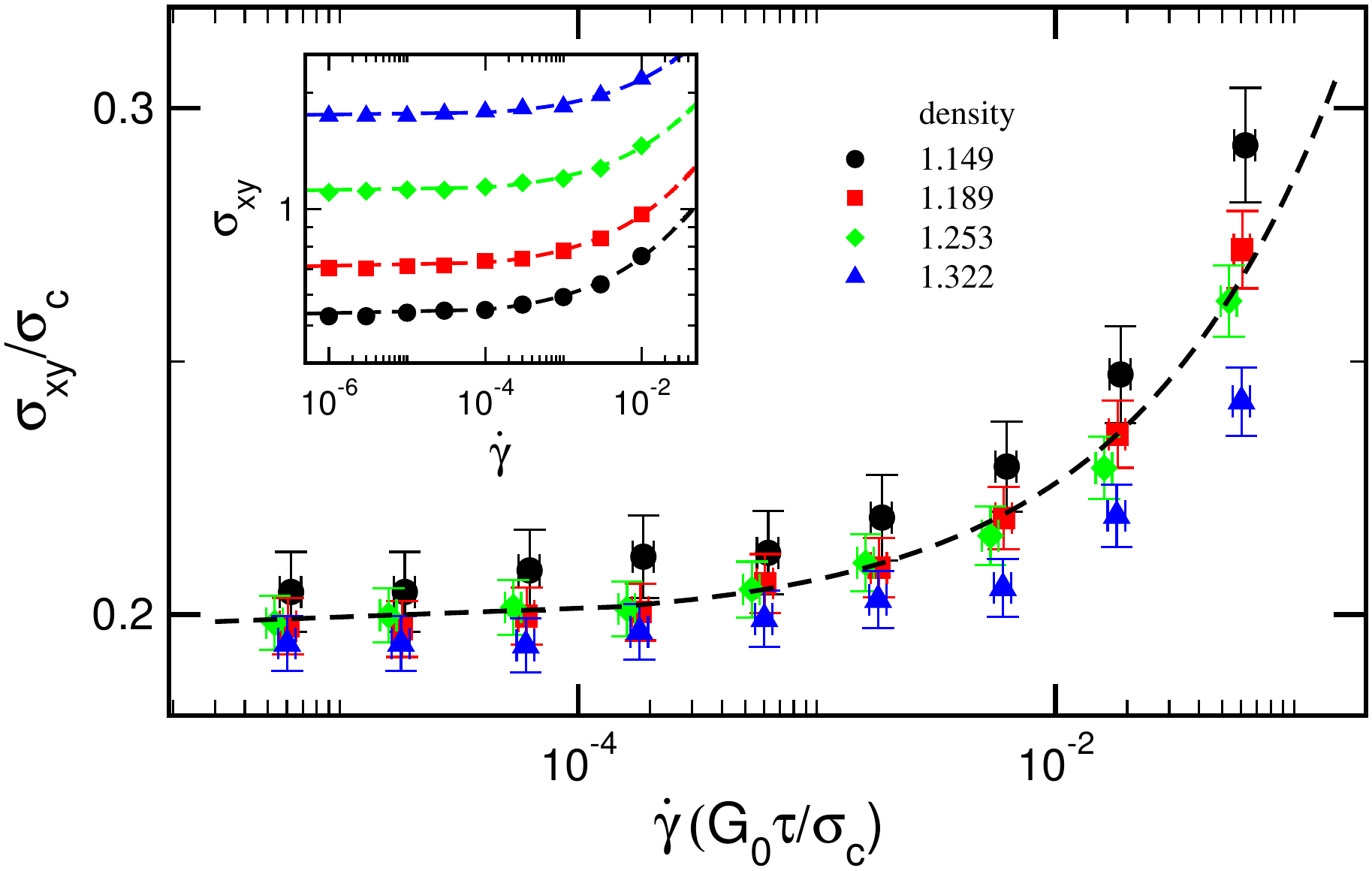}
\end{center}
\caption{{\it Flow curves:} Rescaled macroscopic shear stress $\sigma_{xy}/\sigma_c$ as a function of the rescaled applied shear rate $\dot{\gamma}(G_0\tau_0/\sigma_c)$ for different densities (see text). The dashed line is a fit with the Herschel-Bulkley form $\sigma_{xy}=\sigma_Y+A\dot{\gamma}^{1/2}$. The inset shows the same data before rescaling, dashed lines display a guide to the eye with the same Herschel-Bulkley fitting form.
}
\label{fig:figure2}
\end{figure}

This last relation introduces a non-linearity into Eq.~(\ref{eq:HLmodel}), since the rate of plastic activity itself
depends on the density probability of the mesoscopic stresses. It is this coupling that renders the problem non-trivial and
yields interesting results regarding the behaviour of macroscopic quantities.

This model is known to exhibit a unique stationary state for a finite shear rate in the large time limit,
where the probability density for the mesoscopic stresses
becomes time independent. To determine the time averaged macroscopic stress in the steady 
state one averages over the mesoscopic stresses weighted by the corresponding steady state probability density 
\begin{equation}
 \langle \sigma \rangle=\int \sigma  \mathcal{P}(\sigma)\dd\sigma.
\end{equation}
Using appropriate units we can write the equations in dimensionless quantities, expressing stress related values in units of the local
yield stress $\sigma_c$, time quantities in units of $\tau$, the shear rate in 
units of $ \sigma_c/(G_0\tau)$ and the stress diffusion coefficient in units of $\sigma_c^2/\tau$, leaving only two independent
dimensionless model parameters that determine the flow behaviour, namely the dimensionless shear rate and coupling constant $\tilde{\alpha}$.

\begin{figure}[th]
\begin{center}
\includegraphics[width=\columnwidth, clip]{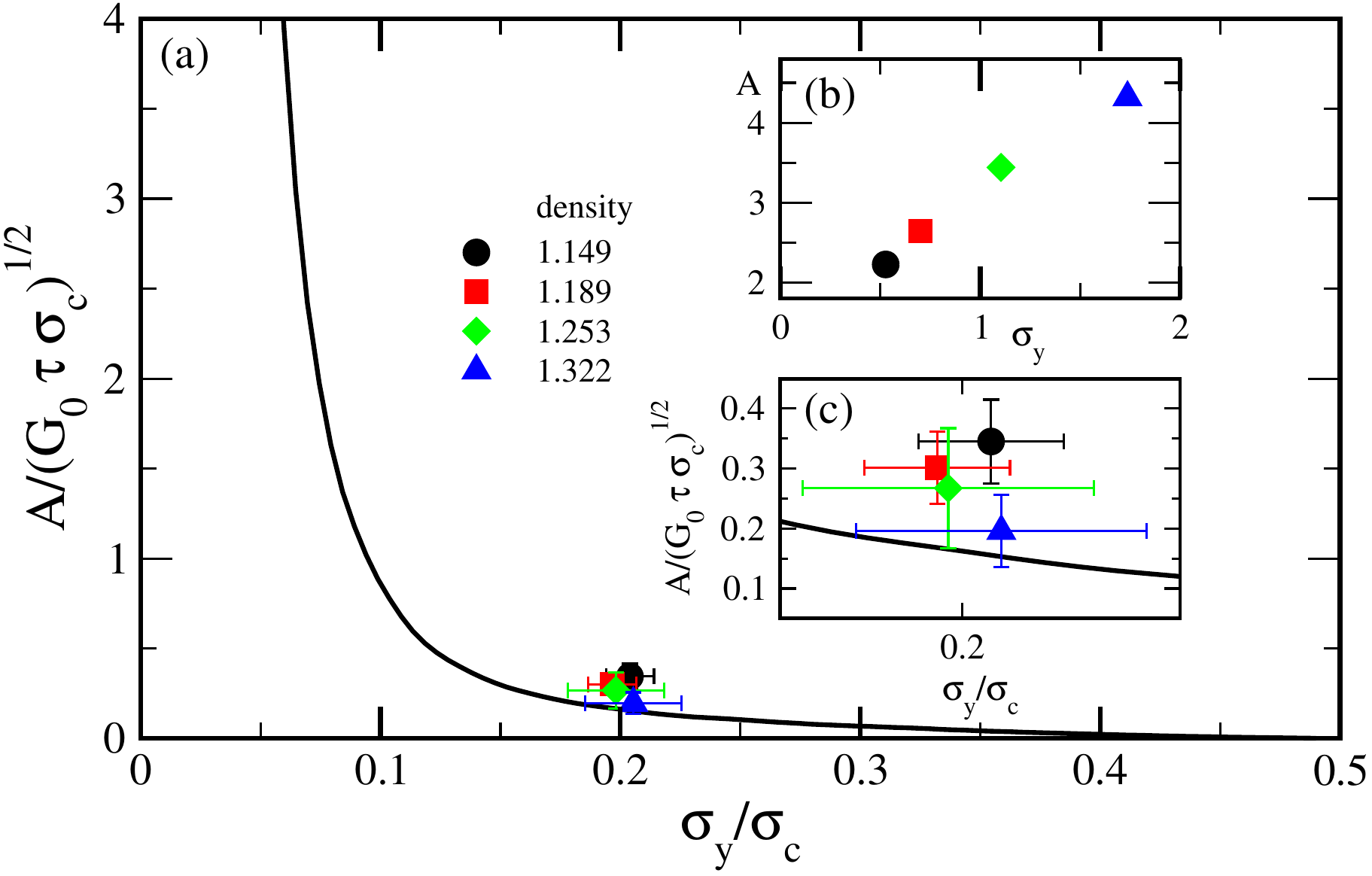}
\end{center}
\caption{{\it HL predictions vs.~MD simulations:} (a) The points are the rescaled Herschel Bulkley prefactor $A$ as a function of the rescaled dynamical yield stress $\sigma_Y$ for four different densities, error bars are estimated from the measurements of the different quantities. The full line corresponds to the HL prediction. (b) This inset shows the data from the simulation fits before the rescaling. (c) This inset shows a zoom into the main panel, to display better the proximity to the theoretical curve obtained from a very simplified picture.
}
\label{fig:figure3}
\end{figure}

The rheological results in the small shear rate limit for this model are well studied \cite{HL, GatiPhD, CancesSIAM06, CancesMMS06, OlivierPhD, OlivierZAMP10, OlivierSIAM11, OlivierSCM12, AgoritsasEPJE15}.  For small enough coupling strength $\tilde{\alpha}<1/2$ 
the HL model predicts a Herschel-Bulkley flow behaviour of exponent $1/2$, $\langle \sigma \rangle\approx \sigma_Y + A\gammap^{1/2}$, for the time averaged macroscopic stress in the steady state $\langle \sigma \rangle=\int \sigma  \mathcal{P}(\sigma)\dd\sigma$,
with the two constants $\sigma_Y$ (the dynamical yield stress) and $A$ (the prefactor).
These macroscopic constants of the model can only depend on the last free control parameter, namely the coupling constant $\tilde{\alpha}$.
This means that we will obtain a universal curve\footnote{For the analytical derivation of this constant relation see the Appendix A} for the rescaled quantities $A/\sqrt{G_0 \tau \sigma_c}$ versus $\sigma_Y/\sigma_c$ parametrized through $\tilde{\alpha}$
(see Fig.~\ref{fig:figure3}).

The physical interpretation of this coupling constant could be a mechanical fragility \cite{HL} that would depend on the details of the microscopic interactions
and thus should be material dependent. In a later work by Bocquet et al., the authors use a spatial approach to the problem that allows for the derivation of an expression for the coupling constant $\tilde{\alpha}$ if one assumes a decorrelation of the plastic event dynamics. 
Within this approximation it is possible to express $\tilde{\alpha}$ as a function of the specific form of the elastic stress propagator \cite{KEP1}.
This suggests that there should be classes of materials with similar elastic responses that not only share the same non-linear exponent
in the flow curve, but also comparable relations between the dynamical yield stress and the prefactor in the flow curve.

In the following we aim at probing not only the above constant relation using microscopic dynamic simulations, but also to test the 
the underlying assumptions, most importantly the self-consistent description of the mechanical noise induced through the plastic activity.
To test the above theory we need to measure all involved parameters that appear in the HL description. 

Quantities and parameters like the local stress and 
the shear modulus $G_0$ are well defined quantities in the microscopic simulations and rather easy to measure. Also, if stress fluctuations
turn out to be diffusive, $D_\text{HL}$ has a well defined meaning. However, other HL parameters like the local yield stress $\sigma_c$, 
the phenomenological parameter $\tau$ and the rate of plastic activity $\Gamma(t)$ are rather difficult to interpret within 
the microscopic picture. In the following we describe our attempt to relate the different parameters to measurable quantities 
in the microscopic dynamics.

\section{Microscopic model}
\label{sec:micromodel}

We have investigated a generic two-dimensional (2D) model of a glass, consisting of a mixture of A and B particles, with $N_A=10400$ and $N_B=5600$, interacting via a Lennard-Jones potential  
$ V_{\alpha\beta}(r)= 4 \epsilon_{\alpha \beta}[ ( {\sigma_{\alpha\beta}}/{r} )^{12} - ( {\sigma_{\alpha\beta}}/{r} )^{6}   ] $
with $\alpha,\beta=A,B$ and $r$ being the distance between two particles. The parameters $\epsilon_{AA}$, $\sigma_{AA}$ and $m_A$ define the units of energy, length and mass; the unit of time is given by $\tau_0=\sigma_{AA}\sqrt{(m_A/\epsilon_{AA})}$. We set $\epsilon_{AA}=1.0 $, $\epsilon_{AB}=1.5 $, $\epsilon_{BB}=0.5 $, $\sigma_{AA}=1.0$, $\sigma_{AB}=0.8$ and $\sigma_{BB}=0.88$ and $m_A=m_B=1$. This choice is known to prevent crystallization in 2D at low temperature  \cite{Kob2d09}. The potential is truncated at $r=r_c=2.5$ for computational convenience and periodic boundary conditions are used. 
The equations of motion are integrated using the velocity Verlet algorithm with a time step $\delta t=0.005$. 
The athermal limit is achieved by thermostating the system at zero temperature via a Langevin thermostat \cite{LangevinThermo} with a damping coefficient $\zeta=1$ which corresponds to a strongly overdamped condition for the dynamics \cite{SalernoPRL12}.
This model, which has been widely studied in different versions \cite{FalkPRE57,LemaitrePRL09,SalernoPRL12,NicolasEPJE14}, is usually considered appropriate for colloidal and other soft glasses. This choice is motived by the purpose of investigating  the general aspects of the rheology of athermal yield stress materials.  

To investigate different athermal flow responses, we explore states with different number density $\rho=(N_A+N_B)/V$
by changing the volume $V$ of the system.  Glassy configurations were prepared by quenching equilibrated configurations at $T=1$ to zero temperature with a fast cooling rate. Simple shear is set at a rate $\dot{\gamma}$ by deforming the box dimensions and remapping the particle positions. The quenching protocol has virtually no effect since we focus on the steady state shear (total imposed deformation $\Delta \gamma > 20\%$).  

In order to characterize the plastic activity of the system we consider the $D^2_{min}$ quantity \cite{FalkPRE57}. For a given particle $i$,  $D^2_{min}$ is defined as the minimum over all possible linear deformation tensors $\mathbf{\epsilon}$ of:
\begin{equation}
D^2(i,t,\delta t)=\sum_j \left [\vec{r}_{ij}(t+\delta t) - \left ( \mathbb{I} + \mathbf{\epsilon} \right)\cdot \vec{r}_{ij}(t) \right]^2
\end{equation}
where the index $j$ runs over all the neighbors of the reference particle $i$ and $\mathbb{I}$ is the identity matrix. We set  the time lag to $\delta t=4$.  This value is a compromise between having a good signal, i.e.~large irreversible displacements, and being able to resolve individual plastic events.

\subsection{Macroscopic flow curve}

In Fig.~\ref{fig:figure2} we show the dependence of the macroscopic shear stress $\sigma_{xy}$ on the applied shear rate $\dot{\gamma}$. The flow curves are well described by the Herschel-Bulkley (HB) law, $\sigma_{xy}=\sigma_Y+A\dot{\gamma}^{n}$, with an exponent $n\approx 0.55$ (not shown here) which seems not to depend on the density. Fixing the exponent $n$ to the value $0.5$, the one predicted by the HL model in the case of $\tilde{\alpha}<1/2$, gives indistinguishable results. Although other works on sheared disordered material in two and three dimensions report similar values for the flow curve exponent \cite{NicolasEPJE14, FuscoEPJE14}, recent works in the literature seem to suggest that a proper finite size scaling analysis close to the yielding transition reveals different critical dynamical exponents \cite{LinPNAS14,BudrikisPRE13,SalernoPRL12,LiuArXiv15}. We would like to insist here on the fact, that our study does not aim at measuring the critical exponents of the transition; instead  we rather test the consistency of the assumptions made in the HL approach in a parameter regime that fits well the model predictions.

\subsection{Size of an elementary plastic region}

\begin{figure}[ht]
\begin{center}
\includegraphics[width=\columnwidth, clip]{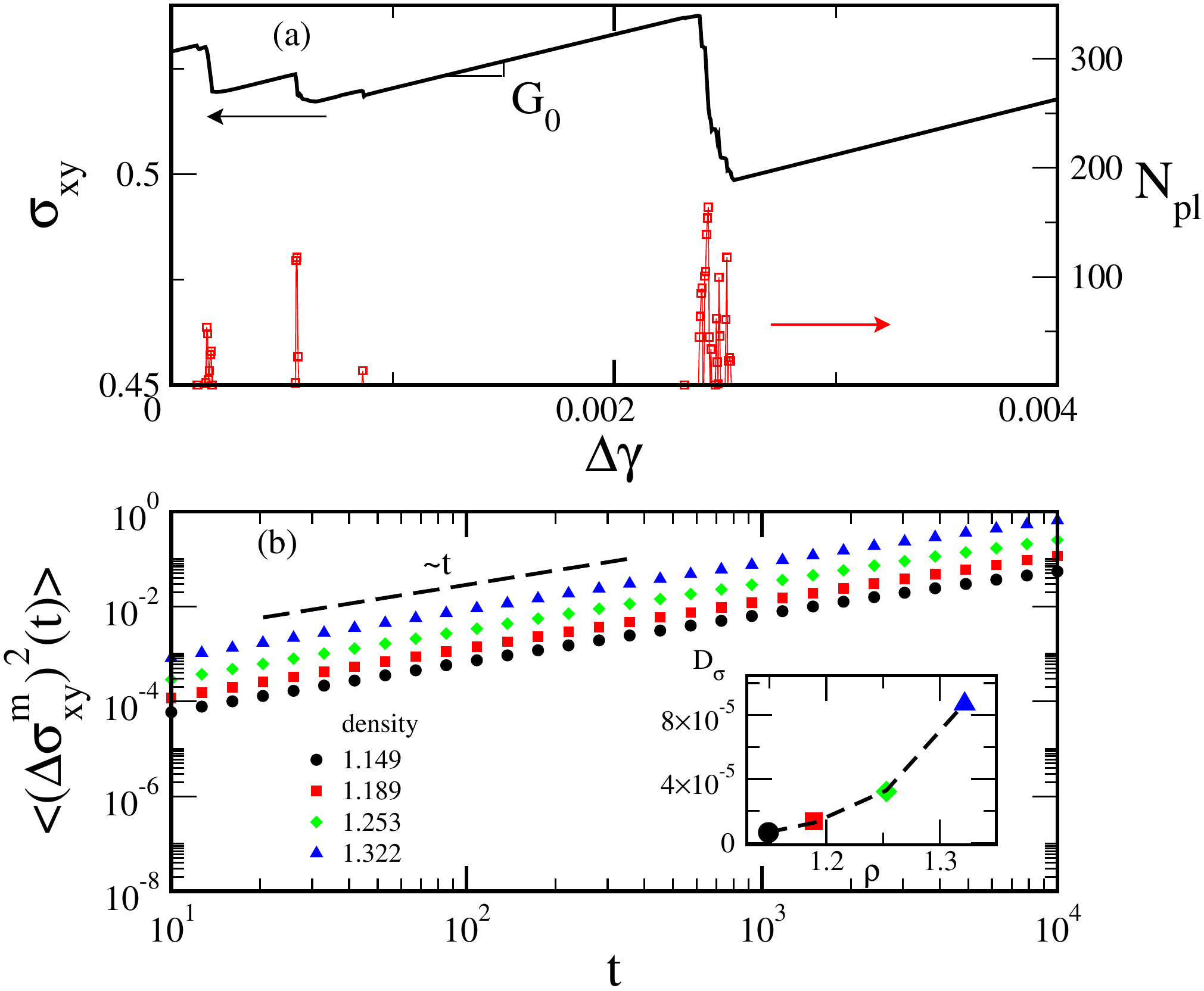}
\end{center}
\caption{{\it Activity and stress diffusion:} (a) The full line represent a part of the macroscopic stress-strain curve in the steady state regime for a density $\rho=1.149$ and a shear rate $\dot{\gamma}=10^{-6}$. The average slope of the elastic parts on the curve yields $G_0$. Open symbols: corresponding evolution of the number of active particles $N_{pl}$ in the system. (b) Coarse mean-square stress difference $\langle ( \Delta \sigma_{xy}^m )^2 (t) \rangle$  as a function of time for different values of the density. Note the linear behaviour at short times. Inset:  stress diffusion coefficient $D_\sigma=\langle ( \Delta \sigma_{xy}^m )^2 (t) \rangle /t $ as a function of density. }
\label{fig:figure4}
\end{figure}

Here we describe the procedure we followed to convert the microscopic simulations into a mesoscopic description. First, we denote a particle $i$ as active, i.e.~performing a plastic rearrangement, at a given time $t$ if the corresponding $D^2_{min}(i,t)$ is larger than a threshold value that we fix equal to $0.1$. In Fig.~\ref{fig:figure4}(a) we compare a typical stress-strain curve and the corresponding evolution of the total number $N_{pl}$ of active particles. The correlation between the stress drops and the peaks in $N_{pl}$ suggests that the actual definition is reasonable.  Then a high-resolution discretization of the system is performed by dividing it into $k\times k$ square blocks of length $w_0=2$. A small block is considered as active if it contains at least one active particle.

The mean size, i.e.~the mean linear extension, of a plastic event $\langle l_{pl} \rangle$ can be estimated by a cluster analysis of the spatial arrangement of the active blocks  in the configurations explored by the system. We employed a modified version of the Hoshen-Kopelman algorithm \cite{HoshenPRB76} in order to account for periodic boundary conditions. If we assume  $\langle l_{pl} \rangle=\langle A_{pl}\rangle^{1/2}$ where $A_{pl}$ is the area of a plastic cluster, we obtain $\langle l_{pl} \rangle \approx 6$, with no relevant dependence on the density and shear rate (in the range $\dot{\gamma}\leqslant 10^{-4}$).
%
The value is in accord with previous works reporting  plastic regions with a size of a few particle diameters \cite{TanguyEPJE06,NicolasEPJE14}. Furthermore, this agreement justifies the criterion we employed to define active particles. Indeed, if we improve the resolution in the analysis of the plastic activity by decreasing the threshold on $D^2_{min}$ by a factor $10$,  while the number of active particles increase by a factor $4$, the mean extension of a plastic event $\langle l_{pl} \rangle$ is reduced by half, suggesting that single particle rearrangements are erroneously taken into account.

Next we implement the coarse-graining of microscopic simulations on the scale of individual plastic events. The simulation box is divided into $M \times M$ square blocks with M chosen in order to have $W\approx \langle l_{pl} \rangle $. The local shear stress $\sigma_{xy}^m$ of a block $m$ is defined as:
\begin{equation}\label{eqn:stressm}
\sigma_{xy}^m=-\frac{1}{W^2}\sum_{i\in m} m v_{i,x}v_{i,y}+\frac{1}{2W^2}\sum_{i\in m}\sum_{j=1}^N\frac{\partial V (r^{ij})}{\partial r^{ij}}\frac{r_x^{ij}r_y^{ij}}{r^{ij}}
\end{equation}
where $ v_{i,x}$ and  $v_{i,y}$ are the $x$ and $y$ components of the velocity of the particle $i$, $r^{ij}$ is the distance between the particles $i$ and $j$ and  the summation of $i$ is performed over the particles in the block. The macroscopic stress tensor $\sigma_{xy}$ is obtained by the summation of $\sigma_{xy}^m$ over all the blocks.

\subsection{Stress diffusion and duration of a plastic event}
The above introduced mean-field model assumes local stress fluctuations, obeying a normal diffusion process in stress space. To test this idea we define the coarse mean-square stress difference as:
\begin{equation}
\langle ( \Delta \sigma_{xy}^m )^2 (t) \rangle = \langle  (\sigma_{xy}^m(t_0+t) - \sigma_{xy}^m(t_0)  -G_0\dot{\gamma}t)^2  \rangle \label{eqn:dsigma}
\end{equation}
where $\sigma_{xy}^m(t)$ is the stress in a given block at time $t$ and the last term in Eq.~(\ref{eqn:dsigma}) accounts for the stress increase due to the elastic deformation of the system, being $G_0$ the macroscopic shear modulus.  In Fig.~\ref{fig:figure4}(b), the short time behavior of $\langle ( \Delta \sigma_{xy}^m )^2 (t) \rangle $ is shown for a finite shear rate $\dot{\gamma}=10^{-6}$, approaching the quasistatic limit.  We observe that at short times, $ \langle ( \Delta \sigma_{xy}^m )^2 (t) \rangle $ increases linearly with time.  We define the stress diffusion coefficient as $D_\sigma=\langle ( \Delta \sigma_{xy}^m )^2 (t) \rangle /t$. In the inset of Fig.~\ref{fig:figure4}(b) we show the dependence of  $D_\sigma$ on the density of the system. 

To define a duration of a plastic event we analyse the two-time autocorrelation function of the $D^2_{min}$ quantity for active particles. In Fig.~\ref{fig:figure5}(a) we show this
correlation function $C_p=\langle D^2_{min}(t_0) D^2_{min}(t_0+t) \rangle / \langle (D^2_{min}(t_0) )^2 \rangle$ as a function of time. We observe that $C_p$ decays exponentially with a characteristic time $\tau_p$, that depends weakly on density (see the inset of Fig.~\ref{fig:figure5}(a)) and that is close to the damping time $\tau_d=\xi^{-1}$ of the Langevin thermostat. We choose to interpret this decorrelation time as the typical duration of a plastic event entering the HL model description.

\subsection{Local yield stress}

In this section, we present a method which allows us to calculate a local critical stress, i.e., the stress limit before a plastic rearrangement occurs locally. 
For this purpose, we adopt the ``frozen matrix" method  \cite{SollichBarra, SollichISGFG10, MizunoPRE13}. The system is frozen except for a target region, i.e., a mesoscopic block, and it's  subjected to a simple shear deformation, with a quasi-static protocol. The frozen region can only deform affinely whereas the target block is allowed to relax non-affinely. For small deformations the target region behaves elastically and the stress increases linearly, with a slope controlled by the local shear modulus  \cite{MizunoPRE13}.  As the strain increases, the elastic behavior goes on until the local yield stress is reached and a plastic rearrangement takes places. This is indicated by a stress drop in the stress-strain curve. For a given block, we define as the local yield stress $\sigma_c^m$  the value of the local stress at the first maximum.  

\begin{figure}[ht]
\begin{center}
\includegraphics[width=\columnwidth, clip]{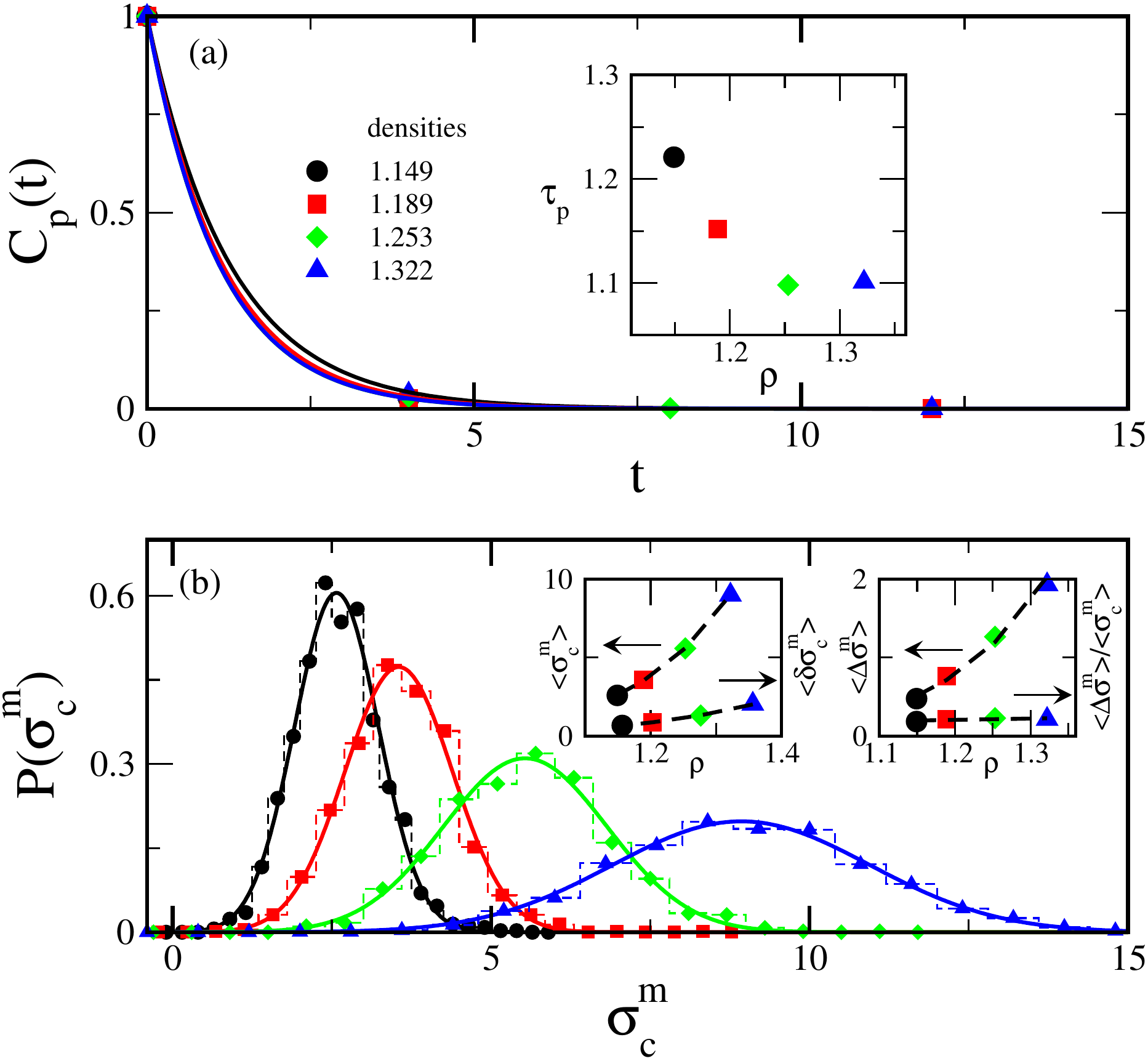}
\end{center}
\caption{{\it Duration of events and local yield stress:} (a) $D^2_{min}$ autocorrelation function for active particles $C_p=\langle D^2_{min}(t_0) D^2_{min}(t_0+t) \rangle / \langle (D^2_{min}(t_0) )^2 \rangle$ as a function of time. Lines are fitting curves $\exp(-t/\tau_p)$. Inset: dependence of $\tau_p$ on the density. (b) Distribution of the local yield stress p($\sigma_Y^m$) for different values of the density. Full lines are Gaussian distributions fits. Left inset: Mean value $\langle \sigma_c^m \rangle$ and variance $\langle \delta \sigma_c^m \rangle$ as a function of density. Dashed lines are power-law guides.  Right inset: Absolute local stress drop $\langle \Delta \sigma^m \rangle$ and the relative local stress drop  $\langle \Delta \sigma^m \rangle / \langle  \sigma_c^m \rangle$ as a function of density. Dashed lines are power-law and constant value guides to the eye. 
Symbol code is the same as in panel (a).}
\label{fig:figure5}
\end{figure}

In Fig.~\ref{fig:figure5}(b) we show the distribution  of  $\sigma_c^m$ for the different values of the density.  First, the fact the local yield stress is distributed is in clear contrast with the HL assumption of a unique critical stress $\sigma_c$. The exposed distributions are well described by a simple Gaussian forms.  The mean values  $\langle \sigma_c^m \rangle$ and the variance  $\langle \delta \sigma_c^m \rangle$ of the distributions are shown in the inset of the figure. As the density increases, $\langle \sigma_c^m \rangle$  increases, due to the enhancement of the repulsive interactions between particles. 

With the frozen matrix method we estimated also the stress release following a local yielding event. 
We observe that the plastic event only partially relaxes the accumulated stress in contrast with the assumption of the HL model of a complete relaxation. The fraction of the relaxed stress seems not to depend on density being $\langle \Delta \sigma^m \rangle / \langle  \sigma_c^m \rangle  \approx 0.2 $ (see inset of Fig.~\ref{fig:figure5}). 

In a former work \cite{MizunoPRE13} it was shown that the mean value of the shear modulus obtained with the frozen matrix method depends on the target region size $W$ and that it converges, from higher values, to the macroscopic modulus as $W$ increases. This is due to the frozen environment which reduces the non-affine motion of the particles in the target region. For $W\approx 6$ the discrepancy is up to $50\%$, a significant error. We are aware that the estimate of $\langle \sigma_c^m \rangle$ may be affected by this effect. Also we have to be careful in the interpretation of the result, because by using the frozen matrix method we measure the yield stress distribution, obtained from strained configurations. It would be interesting to try to infer the inherent yield stress distribution from our measurements \cite{AgoritsasEPJE15}, but this is beyond the scope of this study and left for further investigations.

\section{Robustness of the HL model}

In the following we aim at probing the robustness of the HL model by comparing our data from the microscopic simulations to the various
assumptions and predictions of the mean-field description. 

Let us first recall some of the basic assumptions made to write the evolution Eq.~(\ref{eq:HLmodel}) for the probability
distribution of the mesoscopic stress: 
First (a) the yielding happens at a constant homogeneous yield stress $\sigma_c$.
Then (b) the local response to surrounding plastic events is encompassed through local diffusive stress fluctuations with a well defined
diffusion coefficient $D_{HL}(t)$. And the third assumption (c) concerns the relaxation of the local stress to zero, 
once a site yields, leading to a typical stress jump $\Delta\sigma_{HL}\approx\sigma_c$. 
The HL model assumptions, aiming for a self-consistent description of the mechanical noise, thus predict a diffusion coefficient proportional 
to the square of this typical local stress jump $\Delta\sigma_{HL}$ divided by a typical time scale, given by the inverse plasticity rate $\Gamma(t)$.

(a) Our data analysis reveals that a homogeneous yield stress is of course not verified in a disordered system, where one expects a distribution of yield stresses (see Fig.~\ref{fig:figure5}b).
However, it has been recently shown that the existence of a local yield stress distribution in the HL dynamics does not strongly alter the
predictions for the flow behaviour \cite{AgoritsasEPJE15}.

(b) We tested as well the second assumption of a normal diffusion of the mesoscopic stresses, and we find that within the range of shear rates
that we consider, the measurement of the average mean-square stress differences indeed allows for the determination of a well-defined diffusion coefficient.
The HL coupling constant $\tilde{\alpha}_\text{micro}^{HL} = D_\sigma / (\sigma_c^2 \Gamma)$  (see Eq.~(\ref{eq:DHL}))
with $D_\sigma$ from the results in Fig.~\ref{fig:figure4}(b) and $\Gamma$ from the cluster analysis turns out to be very small, with $\tilde{\alpha}_\text{micro}^{HL}$ of order $10^{-2}$ for all considered densities.

(c) However, our study suggests that the partial relaxation of the local stresses after yielding, $\langle \Delta \sigma^m \rangle / \langle  \sigma_c^m \rangle  \approx 0.2 $ (instead of the assumed total relaxation) introduces an important corrective factor for the coupling constant. Since the typical stress jump in the diffusion process is now reduced, we obtain for the stress diffusion coefficient an altered expression
\begin{equation}\label{eq:DHLcorr}
D_\sigma = \tilde{\alpha}_\text{micro} (\Delta \sigma^m)^2 \Gamma.
\end{equation}
We report the values obtained for this new definition of the coupling constant in Table \ref{tab:testmu}. It turns out that these values are in good approximation density independent and yield an average value of about $\tilde{\alpha}\approx 0.26$. Thus the coupling constant displays a reasonable value smaller than $0.5$ for which the HL model 
predicts \cite{AgoritsasEPJE15}  a well-established Herschel Bulkley regime in the flow curve with an exponent of $1/2$ .

\begin{table*}[t]
\centering
\setlength{\extrarowheight}{1mm}
\begin{tabular}{cc|cc|cc}
 density &&$\tilde{\alpha}_{micro}/10^{-1}$ & & $\tilde{\alpha}_{flow}/10^{-1}$& \\
\hline
$1.149$ &&$2.81\pm 0.21$& \rdelim\}{4}{2cm}[\ $2.57\pm 0.26$]&$3.21\pm 0.12$  &\rdelim\}{4}{2cm}[\ $3.28\pm 0.16$]\\
$1.189$ & &$2.47\pm 0.27$ &&$3.30\pm 0.19$  & \\
$1.253$ & &$2.27\pm 0.24$&&$3.28\pm 0.18$  & \\
$1.322$ & &$2.74\pm 0.30$&&$3.34\pm 0.18$  &\\
\hline
\end{tabular}
\caption{Comparison of different coupling constant measurements as explained in the text.}
\label{tab:testmu}
\end{table*}

And indeed our flow curves from the microscopic simulations can be well fitted with an Herschel Bulkley expression $\sigma_{xy}=\sigma_y+A \dot{\gamma}^{1/2}$ (see inset of Fig.~\ref{fig:figure2}). In Fig.~\ref{fig:figure2} we plot our data in dimensionless units, rescaled as suggested by the HL model, using dimensionless quantities $\sigma_{xy}/\sigma_c$ and $\dot{\gamma}(G_0\tau/\sigma_c)$.
The shear modulus $G_0$ can be easily accessed in simulations as the slope of the stress-strain curve in the elastic regime. We approximate the critical stress $\sigma_c$ as the mean value $\langle \sigma_c^m \rangle$ of the local yield stress obtained with the frozen matrix method and we associate the phenomenological parameter $\tau$  with the duration of a plastic event in the low shear rate limit \cite{AgoritsasEPJE15}. Hence, we will approximate $\tau$ by the value of $\tau_p$ obtained through the measurement of the $D^2_{min}$ two-time autocorrelation function. We observe a collapse of the flow curves for different densities. This collapse is not perfect, but regarding the large error bars introduced by our methods and estimations it appears still convincing and suggest a very generic flow behaviour. 

One of the strongest prediction of the HL model is that both the prefactor $A$ as well as the dynamical yield stress $\sigma_y$ are determined solely by the specific value of the coupling constant between diffusion and plastic activity. 
In Fig.~\ref{fig:figure3} we compare the dimensionless yield stress $\sigma_y/\langle \sigma_c^m \rangle$ and prefactor $A/(G_0\sigma_c\tau)^{1/2}$, obtained from a Herschel Bulkley fit of the flow curves, with the analytically obtained parametric curve. We observe that all the data belonging to different density values collapse as expected by the previous collapse of the data roughly onto a single point close to the theoretical curve, that lies within the estimated error. This result points to a universal determination of the flow curves, a priori strongly sensitive to the density (note the density dependence of $A$ and $\sigma_y$ in the inset of Fig.~\ref{fig:figure3}), through one single density independent parameter, namely the coupling constant $\tilde{\alpha}$. As suggested by the KEP model \cite{KEP1}, we expect this parameter to be only dependent on the specific form of the elastic propagator \cite{PuosiPRE14}. 

If we assume the HL parametric relation between prefactor $A(\tilde{\alpha})$ and dynamical yield stress $\sigma_y(\tilde{\alpha})$ to hold, it is possible to estimate
the coupling constant $\tilde{\alpha}$ in an alternative way through the macroscopic measure of the flow curve, in the following referred to as $\tilde{\alpha}_\text{flow}$.
Despite all the rough approximations we had to make, that tend to introduce large error bars on the data, we find the comparison between the coupling constant $\tilde{\alpha}_{micro}$ with the alternative measurement using the expression for the rescaled yield stress\footnote{given in Appendix A} $\sigma_y/\sigma_c(\tilde{\alpha})$ rather convincing (see table \ref{tab:testmu}). Both measurements suggest a density independent result with a quite small relative error of approximately $20 \%$.

Altogether our data suggests that revisiting the rule of setting the stress to zero after a yield event in the HL model equations, by changing the gain term in the evolution Eq.~(\ref{eq:HLmodel}) together with the introduction of a more realistic yield stress distribution, seems to be a promising route to reach a more realistic mean-field modeling of athermally sheared amorphous systems. 

\section{Conclusion}

In this study we aimed at testing some of the most basic assumptions and predictions of mean-field modeling for the rheology of athermally sheared amorphous systems. In conclusion we obtain a consistent picture of how to model correctly the mechanical noise in the regime of large enough driving rates (far from the true critical point \cite{LiuArXiv15}). We find that we can incorporate the noise into a normal diffusion of local stresses with a noise amplitude solely governed by the rate of plastic activity as proposed by the H\'ebraud-Lequeux (HL) model \cite{HL}. We not only confirm this physical picture using molecular dynamics simulations, but we also show that the coupling strength between diffusion
and the rate of plastic activity, a dimensionless and density independent quantity, seems to determine the specific form of the rheological response. 
Our data analysis suggest some important modifications in the original version of HL model equations such as the partial relaxation of the local stress after a yielding event and the introduction of a yield stress distribution obtained from the microscopic simulations. In a future work we plan to test the coherence of such a modified model with our microscopic approach. Further it would be highly desirable to test the degree of generality of the results on other model systems \cite{FuscoEPJE14} and experimental setups, that have access to the measure of local plastic activity \cite{AmonPRL12, DesmondSoftMat13, KnowltonSoftMat14, LeBouilPRL14, JensenPRE14}.

{\it Acknowledgments}
KM acknowledges financial support of the French Agence Nationale de la Recherche (ANR), under grant ANR-14-CE32-0005 (project FAPRES) and FP acknowledges financial support from ERC grant ADG20110209. Most of the computations were performed using the Froggy platform of the 
\href{https://ciment.ujf-grenoble.fr}{CIMENT} infrastructure supported by the Rh\^one-Alpes region (GRANT CPER07-13
\href{http://www.ci-ra.org/}{CIRA}) and the Equip@Meso project (reference ANR-10-EQPX-29-01).
Further we would like to thank Elisabeth Agoritsas, Eric Bertin and Jean-Louis Barrat for fruitful discussions and valuable revisions of our manuscript. Also we would like to thank J\"org Rottler for assistance
in the cluster analysis of plastic events.

\begin{appendix}
\section*{Appendix A}

\subsection*{The parametric relation between the yield stress and the HB prefactor}
\label{app:calculus}

In this section, we compute the relation between the dynamic yield 
stress $\sigma_Y$ and the prefactor $A$ in the relation 
$\langle\sigma\rangle=\sigma_Y+A \dot{\gamma}^{1/2}$ derived from the HL 
model (as given in Fig.~3). We use the method developed in \cite{OlivierPhD}: the probability density 
function solving the (stationary) HL equation (2) of the paper is 
expanded in the following way
\begin{align} 
\mathcal{P}(\sigma)&=Q^0(\sigma)+\dot{\gamma}^{1/2}Q^1(\sigma)+\cdots,
&\text { for $\sigma$ in $[-\sigma_c,\sigma_c]$}\label{eq:ansatzq}\\ 
\mathcal{P}(\sigma)&=\dot{\gamma}^{1/2}R^1_{\pm}
\left(\frac{|\sigma|-\sigma_c}{\dot{\gamma}^{1/2}}\right)+\cdots , &\text{ 
for 
$\pm\sigma$ in $[\sigma_c,+\infty[$}\label{eq:ansatzr}
\end{align}
This ansatz has been proved to be correct in the case where 
$\alpha<\sigma_c^2/2$ (again see \cite{OlivierPhD}). Moreover, term-by-term 
integration is allowed which means that
\begin{multline}
\langle\sigma\rangle=\int\sigma\mathcal{P}
(\sigma)\dd\sigma=\underbrace{\left(\int \sigma 
Q^0(\sigma)\dd\sigma\right)}_{\sigma_Y}\\+\underbrace{\left(\int \sigma 
Q^1(\sigma)\dd \sigma\right)}_{A}\dot{\gamma}^{1/2}+\cdots
\end{multline}
where the dots are terms of higher order than $\dot{\gamma}^{1/2}$. Now 
all that is left to do is to identify $Q^0$ and $Q^1$. By plugging 
\eqref{eq:ansatzq} in \eqref{eq:HLmodel} we obtain the following equation 
on $Q^0$:
\begin{equation}
\begin{cases}
 -d\d_{\sigma}^2Q^0+\tau G_0\d_\sigma Q^0=\frac{d}{\alpha}\delta_0\\
 Q^0(\pm\sigma_c)=0\\
 \int_{-\sigma_c}^{\sigma_c} Q^0(\sigma)\dd \sigma=1
 \end{cases}
\end{equation}
Note that in this equation $d$ is an unknown coefficient used to enforce 
the integral condition on $Q^0$; it is physically related to the diffusion 
coefficient by $\tau D\sim d\dot\gamma$.

This system can easily be integrated and we find $Q^0$ to be

\begin{equation}\label{eq:q0}
Q^0(\sigma)=\frac{1}{\sigma_c(1-e^{\frac{-\sigma_cG_0\tau}{d}})}
\begin{cases}
             \left(e^{\frac{\sigma G_0\tau}{d}}
-e^{\frac{-\sigma_c\tau 
G_0}{d}}\right)&-\sigma_c\leq\sigma\leq 0\\
1-e^{\frac{(\sigma-\sigma_c)G_0\tau}{d}}&0\leq\sigma\leq\sigma_c
            \end{cases}
\end{equation}
and $d$ is selected so that the following equation holds true:
\begin{equation}
 \frac{\alpha}{\sigma_c^2}=\frac{d}{\sigma_c 
G_0\tau}\tanh\left(\frac{\sigma_c G_0\tau}{2d}\right)
\end{equation}
Note that this equation (in $d$) has a unique solution $d(\alpha)$ if and 
only if $0<\alpha/\sigma_c^2<1/2$. However, in view of \eqref{eq:q0}, it 
is easier to express everything in terms of the parameter $d$ instead of 
$\alpha$: then the limit $\alpha/\sigma_c^2\to 1/2$ is equivalent to $d\to 
+\infty$ and $\alpha\to 0$ is equivalent to $d\to 0$.

Now using the formula for computing $\sigma_Y$ we obtain by integration,
\begin{equation}
\frac{\sigma_Y}{\sigma_c}=\frac{1}{2}\coth\left(\frac{\sigma_cG_0\tau}
{2d} \right)-\frac{d}{\sigma_c G_0\tau}
\end{equation}

To obtain the prefactor $A$ we must compute $Q^1$ which necessitates the 
computation of $R^1_\pm$. Using the continuity of 
$\d_\sigma\mathcal P$ at $\pm\sigma_c$, one can find that $R^1_\pm(z)$ are 
functions satisfying

\begin{equation}
 \begin{cases}
  -d\d_{z}^2R^1_\pm+R^1_\pm=0\\
  \d_zR^1_+(0)=\d_\sigma 
Q^0(\sigma_c)=-\frac{\tau G_0 
}{d\sigma_c(1-e^{\frac{-\sigma_c\tau 
G_0}{d}})}\\
  -\d_zR^1_-(0)=\d_\sigma 
Q^0(-\sigma_c)=\frac{\tau G_0e^{\frac{-\sigma_c\tau 
G_0}{d}}}{d\sigma_c(1-e^{-\frac{\sigma_c\tau G_0}{d}})}
 \end{cases}
\end{equation}

Again, it is easy to solve this system:
\begin{align}
 R^1_+(z)&=\frac{\tau G_0 
}{\sqrt{d}\sigma_c(1-e^{\frac{-\sigma_c\tau 
G_0}{d}})}e^{\frac{-z}{\sqrt{d}}}\\
R^1_-(z)&=\frac{\tau G_0e^{\frac{-\sigma_c\tau 
G_0}{d}}}{\sqrt{d}\sigma_c(1-e^{-\frac{\sigma_c\tau 
G_0}{d}})}e^{\frac{-z}{\sqrt{d}}}
\end{align}

Now we can write down the equations satisfied by $Q^1$ using the 
continuity of $\mathcal P$ at $\sigma=\pm\sigma_c$:
\begin{equation}
\begin{cases}
 -d\d_\sigma^2Q^1+G_0\tau \d_\sigma Q^1=\tilde 
d(\frac{1}{\alpha}\delta_0+\d_\sigma^2Q^0)\\
Q^1(\sigma_c)=R^1_+(0)=\frac{\tau G_0 
}{\sqrt{d}\sigma_c(1-e^{\frac{-\sigma_c\tau 
G_0}{d}})}\\
Q^1(-\sigma_c)=R^1_-(0)=\frac{\tau G_0e^{\frac{-\sigma_c\tau 
G_0}{d}}}{\sqrt{d}\sigma_c(1-e^{-\frac{\sigma_c\tau 
G_0}{d}})}\\
\int_{-\sigma_c}^{\sigma_c} Q^1(\sigma)\dd\sigma=0
\end{cases}
\end{equation}
In this system $\tilde d$ is an unknown coefficient to be 
simultaneously computed  with $Q^1$. The computation of $Q^1$ is tedious 
but straightforward. The expression of $Q^1$ are quite lengthy but can be 
checked out on a symbolic computation program: if $-\sigma_c\leq\sigma\leq 
0$ then
\begin{equation}
\begin{split}
 Q^1(\sigma)&=\frac{\tau G_0}{\sigma_c\sqrt{d} \left( 
1-e^{ \frac{-G_0  \tau \sigma_c}{d}}\right)  
} e^{ \frac{G_0 \tau}{d} \sigma} 
\\
&\quad -\frac{ \tilde d G_0 \tau}{d^2\sigma_c \left( 
1-e^{ \frac{-G_0  \tau \sigma_c}{d} }\right) } 
\left( \sigma+\sigma_c\right) 
 e^{ \frac{G_0 \tau}{d} \sigma}\\
 &\quad +\frac{1}{e^{ 
\frac{G_0 \tau}{d} \sigma_c} 
-e^{ \frac{-G_0\tau}{d} \sigma_c} }\times\\
&\quad\quad
\left(\frac{\sigma_c G_0  \tau}{d} \left( 
\frac{ \tilde d}{d\sigma_c}-\frac{\sqrt{d}}{\sigma_c^2}\right) 
+\frac{\tilde d 
}{d\sigma_c} \left( 
1+e^{ \frac{-G_0  \tau}{d} \sigma_c }\right) 
\right) \times\\
&\quad\quad\left( 
e^{ \frac{G_0 \tau}{d} \left(\sigma+\sigma_c\right)}-1\right)
\end{split}
\end{equation}
and if $0\leq\sigma\leq \sigma_c$
\begin{equation}
\begin{split}
 Q^1(\sigma)&=\frac{\tau G_0}{\sigma_c\sqrt{d} \left( 1-e^{
-\frac{G_0  \tau \sigma_c}{d}} \right) 
}  e^{ \frac{G_0 \tau}{d} \left( \sigma-\sigma_c\right)} 
\\
&\quad+\frac{\tilde d G_0 \tau}{d^2\sigma_c 
\left( 1- e^{-\frac{G_0  \tau \sigma_c}{d}} 
\right) } \left( \sigma-\sigma_c\right)  e^{
\frac{G_0 \tau}{d} \left( \sigma-\sigma_c\right)}
\\
&\quad+\frac{1}{ e^{ \frac{G_0 \tau}{d} \sigma_c} 
- e^{-\frac{G_0  \tau}{d} \sigma_c} 
}\times\\
&\quad\quad\left(-\frac{\sigma_c G_0  \tau}{d} \left( 
\frac{\tilde d}{d\sigma_c}-\frac{\sqrt{d}}{\sigma_c^2}\right) +\frac{ 
\tilde d}{d\sigma_c} 
\left( 1+ e^{\frac{G_0  \tau}{d} \sigma_c} 
\right) \right)\times\\
&\quad\quad\left( 1-e^{\frac{G_0 \tau}{d} \left( 
\sigma-\sigma_c\right)} \right)
\end{split}
\end{equation}
The parameter $\tilde d$ is selected to enforce the vanishing integral 
condition which amounts to taking
\begin{equation}
 \tilde d=-\frac{d^{3/2}}{\sigma_c}\frac{e^{\frac{\tau 
G_0\sigma_c}{d}}-e^{-\frac{\tau 
G_0\sigma_c}{d}}+\frac{2\tau G_0\sigma_c}{d}}{e^{\frac{\tau
G_0\sigma_c}{d}}-e^{-\frac{\tau 
G_0\sigma_c}{d}}-\frac{2\tau G_0\sigma_c}{d}}
\end{equation}

Finally we can compute the prefactor $A$ which is equal to $\int \sigma 
Q^1(\sigma)\dd \sigma$. All in all, we obtain (let us note 
$u=(\tau G_0\sigma_c)/d$):
\begin{align}
 \frac{\sigma_Y}{\sigma_c}&=\frac{1}{2}\coth\left(u \right)-\frac1u\\
\frac{A}{\sqrt{\sigma_c \tau 
G_0}}&=\frac{1}{2\sqrt{u}}\frac{e^{{u}}\left({u}
\cosh\left({2u}\right)+2\sinh\left({2u}\right)+{u}\right)}
{{u}\left(\sinh\left({2u}\right)-{u}\right)}\\
&\nonumber\quad-\frac{1}{2\sqrt{u}}\frac{{6 u^2}\coth\left({
2u}\right)}{{u}\left(\sinh\left({2u}\right)-{u}
\right)}
\end{align}

\end{appendix}

\bibliography{biblio_HL}
\bibliographystyle{rsc} 

\end{document}